\documentclass[aps,prb,twocolumn,superscriptaddress,showpacs]{revtex4-1}

\usepackage{amsmath}
\usepackage{graphicx}
\usepackage{calc}
\usepackage{bm}

\usepackage[T2A]{fontenc}
\usepackage[cp1251]{inputenc}
\usepackage[english]{babel}

\begin{document}

\title{Evidence for surface spin structures from first order reversal curves in Co$_3$Sn$_2$S$_2$ and Fe$_3$GeTe$_2$ magnetic topological semimetals}

\author{A.A. Avakyants}
\author{N.N. Orlova}
\author{A.V. Timonina}
\author{N.N. Kolesnikov}
\author{E.V. Deviatov}

\affiliation{Institute of Solid State Physics of the Russian Academy of Sciences, Chernogolovka, Moscow District, 2 Academician Ossipyan str., 142432 Russia}

\date{\today}

\begin{abstract}
We study   magnetization reversal and first order reversal curves for two different magnetic topological semimetals, Co$_3$Sn$_2$S$_2$ and Fe$_3$GeTe$_2$, in a wide temperature range. For the magnetization reversal, we observe strong temperature dependence of the initial (low-temperature) step-like magnetization switchings, so the inverted hysteresis appears at high temperatures. Usually, the inverted hysteresis is a fingerprint of material with two independent magnetic phases, the inversion reflects the phase interaction. First order reversal curve analysis confirms the  two-phase behavior even at the lowest temperatures of the experiment. While the bulk ferromagnetic magnetization shows strong temperature dependence, one of the observed phases demonstrates perfect stability below the  Curie temperature. The  obtained hysteresis loops are of the bow-tie type, which is usually ascribed to appearance of the skyrmionic phase. The described two-phase behavior is mostly identical for Co$_3$Sn$_2$S$_2$ and Fe$_3$GeTe$_2$ magnetic topological semimetals, only the characteristic temperatures differ for    these materials. The specifics of our experiment is the excellent temperature stability of the second phase, while the skyrmions are usually observed near the Curie temperature. On the other hand, temperature stability can be expected for surface-state induced spin textures due to the topological protection of surface states in topological semimetals.   This also explains the universal behavior of the second phase for two different topological semimetals Co$_3$Sn$_2$S$_2$ and FGT. Both these materials have strongly  different bulk properties, the only similarity is the presence of the topological surface states. Thus, we can ascribe the second, temperature-stable magnetic phase to the surface states in topological semimetals.
\end{abstract}


\maketitle

\section{Introduction}

Recent interest to topological semimetals is mostly connected with Fermi arc surface states, which are known for Dirac, Weyl, and nodal-line semimetals~\cite{armitage}. All of them are characterized by  band touching in some distinct nodes (or along a nodal line), which are the special points of Brillouin zone with three dimensional linear dispersion. Topologically protected Fermi arc surface states  are  connecting projections of these nodes on the surface Brillouin zone. 

Most of experimentally investigated WSMs, were non-centrosymmetric crystals with broken inversion symmetry~\cite{armitage}.  In contrast, there are only a few candidates of magnetically ordered materials for the realization of WSMs~\cite{mag1,mag2,mag3,mag4,kagome,kagome1}. Recently,  giant anomalous Hall effect was reported~\cite{kagome,kagome1} for the kagome-lattice ferromagnet Co$_3$Sn$_2$S$_2$, as an indication for the existence of a magnetic Weyl phase~\cite{armitage}. Fermi arcs were directly visualized for Co$_3$Sn$_2$S$_2$  by scanning tunneling spectroscopy~\cite{kagome_arcs}. Also, three-dimensional Fe$_3$GeTe$_2$ (FGT) is a unique candidate for the ferromagnetic nodal-line semimetal~\cite{kim}, hosting spin-polarized Fermi arc surface states~\cite{asymmr}.

Intriguing spin properties of Weyl semimetals make it attractive material for  spin investigations.  Complex spin textures  appear~\cite{jiang15,rhodes15,wang16} in WTe$_2$ due to the strong spin and momentum correlation~\cite{Sp-m-lock} (spin-momentum locking) in topological semimetals.  Spin- and angle- resolved photoemission spectroscopy  data indeed demonstrate  surface Fermi arcs with nearly full spin polarization~\cite{das16,feng2016,lv2015,xu16}. Surface topological textures (skyrmions) are also visualized in some magnetic semimetals by STM, Lorenz electron microscopy, and magnetic force microscopy~\cite{CrGeTe,FGT_skyrmion1,FGT_skyrmion2}. Recent investigations show  topological protection  of skyrmion structures due to their origin from the spin-polarized topological  surface states~\cite{Araki}.

Thus, any magnetic topological semimetal consists of two correlated  spin-ordered systems, which are the magnetically ordered bulk and the spin-polarized surface states. On the other hand, similar two-component magnetic systems are known for the artificial objects like ferromagnetic multilayers, e.g. for Co/Pd~\cite{Co/Pd}, Pt/Co/Ta~\cite{Pt/Co/Ta,Pt/Co/Ta2}, and  Ir/Fe/Co/Pt~\cite{Ir/Fe/Co/Pt} multilayer films. The  nonmagnetic layers are of strong spin-orbit coupling, so magnetic skyrmions appear due to broken inversion symmetry at the interface with the ferromagnetic layers. Multilayers demonstrate two independent magnetization processes~\cite{SrRuOheterostr}, inverted hysteresis~\cite{invhyst}, exchange bias~\cite{exchbias} and spin-valve~\cite{spin valve1,spin valve2} effects. The magnetic textures are also responsible for the atypical magnetization dynamics, i.e. for the bow-tie magnetization hysteresis loops~\cite{bow-tie} in these systems.

Due to the obvious similarity with the ferromagnetic multilayers, it is reasonable to study magnetization dynamics in magnetic  topological semimetals. Recently, current-induced spin dynamics has been investigated in Weyl topological surface states~\cite{timnal,wteni,cosns}, demonstrating strong similarity with spin-polarized transport in multilayers~\cite{myers,tsoi1,tsoi2,katine}. On the other hand, direct demonstration is still missing for the independent magnetization of bulk and surface magnetic systems for topological semimetals.

 Here, we study   magnetization reversal and first order reversal curves for two different magnetic topological semimetals, Co$_3$Sn$_2$S$_2$ and Fe$_3$GeTe$_2$, in a wide temperature range. For the magnetization hysteresis loops, we observe strong temperature dependence of the initial (low-temperature) step-like magnetization switchings, so the inverted hysteresis appears. Usually, similar behavior is ascribed to the second, temperature-induced magnetic phase. However, first order reversal curve analysis confirms the  two-phase behavior even at the lowest temperatures of the experiment. While the bulk ferromagnetic magnetization is expected to have strong temperature dependence, one of the observed phases demonstrates perfect stability below the  Curie temperature. The described two-phase behavior is mostly identical for Co$_3$Sn$_2$S$_2$ and Fe$_3$GeTe$_2$ magnetic topological semimetals, only the characteristic temperatures differ for    these materials.

\section{Samples and techniques}

 Co$_3$Sn$_2$S$_2$ single crystals were grown by the gradient freezing method. Initial load of high-purity elements taken in stoichiometric ratio was slowly heated up to $920^{\circ}$C in the horizontally positioned evacuated silica ampule, held for 20 h and then cooled with the furnace to the ambient temperature at the rate of 20 degree/h. The obtained ingot was cleaved in the middle part. The Laue patterns confirm the hexagonal structure with (0001) as cleavage plane. 
 Fe$_3$GeTe$_2$ single crystals were grown by the two-stage iodine transport from the initially synthesized Fe$_3$GeTe$_2$ compound~\cite{fgtoleg}.
 The electron probe microanalysis of cleaved surfaces and X-ray diffractometry of powdered samples confirmed the stoichiometric composition of the crystals.

To confirm Co$_3$Sn$_2$S$_2$ and Fe$_3$GeTe$_2$ crystals  quality, magnetoresistance measurements were also performed  in standard Hall bar geometry for the reference samples with normal (Au) leads~\cite{cosns,fgtoleg}.   The specific feature of time reversal symmetry breaking in topological semimetals is  a large anomalous Hall effect (AHE), which manifests itself as non-zero Hall conductance in zero magnetic field~\cite{ahe1,ahe2,ahe3}. AHE can be understood in a topological-insulator-multilayer model, where the two-dimensional Chern edge states form the three-dimensional surface states~\cite{armitage}.  We have demonstrated a large anomalous Hall effect for the reference samples~\cite{cosns,fgtoleg}, which indicates   a magnetic topological phase~\cite{armitage}.

To investigate magnetic properties of small Co$_3$Sn$_2$S$_2$ and Fe$_3$GeTe$_2$ single crystal flakes, we  use Lake Shore Cryotronics 8604 VSM magnetometer. It is equipped with nitrogen flow cryostat, which allows measurements below   Curie temperatures of  Co$_3$Sn$_2$S$_2$ and Fe$_3$GeTe$_2$ (177~K and 220~K, respectively)~\cite{Tc}. 

Due to the layered structure of Co$_3$Sn$_2$S$_2$ and Fe$_3$GeTe$_2$ single crystals, small flakes can be easily obtained by  mechanical exfoliation. Afterward, the flake is mounted to the magnetometer sample holder by   low temperature grease, which has been tested to have a negligible magnetic response. The sample holder can be rotated in magnetic field, also, we perform centering and saddling procedures to establish the correct sample position in the cryostat.

We investigate sample magnetization by the standard method of the magnetic field gradual sweeping between two opposite saturation values to obtain hysteresis loops at different temperatures. Apart from the  hysteresis measurements, we perform first order reversal curve (FORC) analysis, which is of growing popularity nowadays.  FORC provides information on the magnetization reversal, which can not be obtained from standard  hysteresis loops. 

The FORC measurements consist of multiple magnetization $M$ recording as a two-dimensional map with the reversal field $H_r$ and the  demagnetization field $H$ as x and y  coordinates, respectively~\cite{FORCanalysis,Hr}. Before every FORC curve, the magnetization is stabilized at some fixed positive saturation field $H_s$. As a second step, the field is changed to the chosen reversal field $H_r$, so the $M(H)$ curve $M(H)$ can be recorded toward the positive magnetic fields. For the next FORC curve, the starting point $H_r$ is shifted to the lower magnetic field, so the FORC density $\rho(H,H_r)$ can be calculated as a second derivative 

\begin{equation}\label{rho}
\rho = -\frac{1}{2} \frac{\partial^2 M(H,H_r)}{\partial H \partial H_r}.
\end{equation}

The obtained $\rho(H,H_r)$ map is usually redrawn in $(H_u,H_c)$ coordinates, where  $H_u = \frac{1}{2}(H + H_r)$ is known as the interaction field and $H_c = \frac{1}{2}(H - H_r)$ is the coercive field. The FORC density distribution $\rho(H_u,H_c)$   is known to be convenient for analysis~\cite{FORCtheory,FORCtheory1}. For example, closed contours are usually associated with single-domain regime, while multi-domain material gives open contours that diverge towards the $H_u$ axis.  Which is of primary importance for us, multiple FORC density $\rho(H_u,H_c)$ structures correspond to multiple magnetic phases in the material~\cite{FORCtheory,FORCtheory1}. 
 In our experiment, the FORC density distribution $\rho(H_u,H_c)$   is calculated by standard Lake Shore Cryotronics software, which is delivered with  Lake Shore Cryotronics 8604 VSM magnetometer. The smoothing factor is 1 for  Co$_3$Sn$_2$S$_2$, while it equals to 2 for Fe$_3$GeTe$_2$.

\section{Experimental results}

\begin{figure}
\includegraphics[width=1\columnwidth]{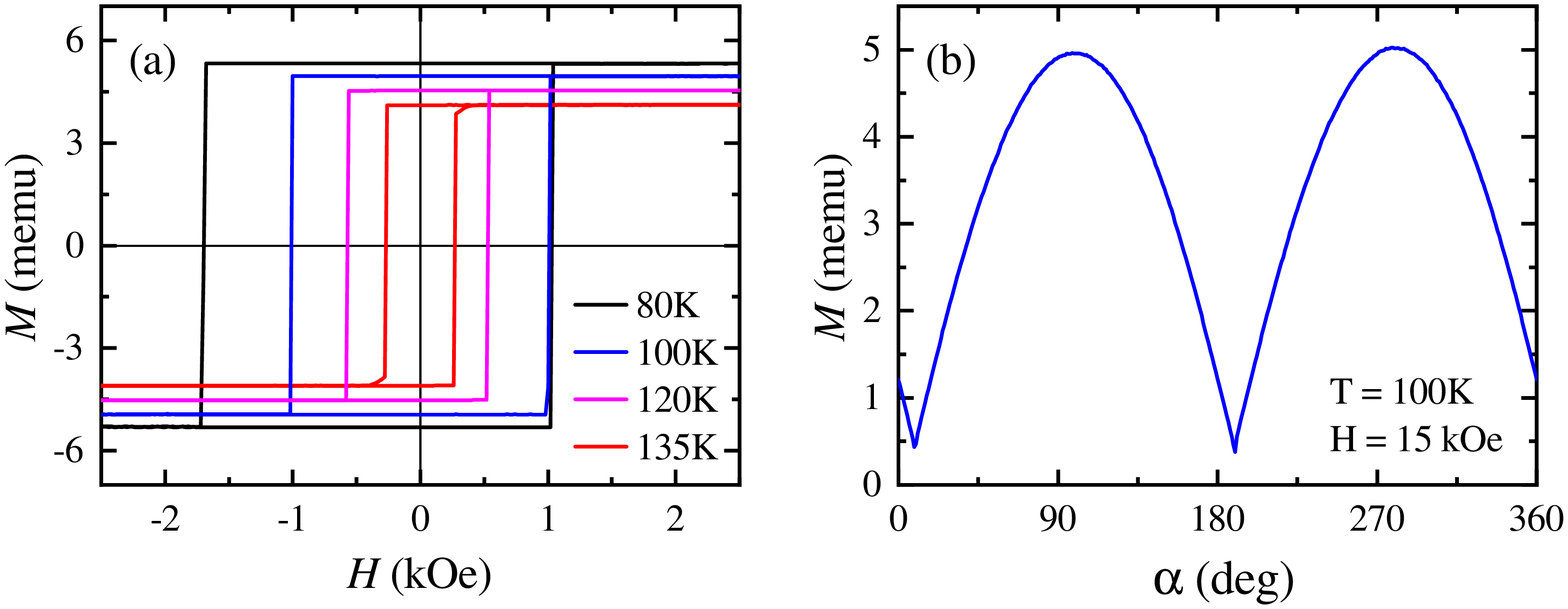}
\caption{(Color online) (a) Rectangular hysteresis loops with step-like magnetization switchings for the   6~$\mu$m thick (0.56~mg) Co$_3$Sn$_2$S$_2$ flake at low temperatures (80~K, 100~K, 120~K and 135~K). The curves confirm the single domain magnetic structure of Co$_3$Sn$_2$S$_2$, the coercive field and the saturated magnetization value are monotonically diminishing with temperature. Here and below, the magnetic field is normal to the flake's surface,  i.e. along the Co$_3$Sn$_2$S$_2$ easy axis direction. (b) Angle dependence of magnetization at $T=100$~K and $H=15$~kOe.  }
\label{single-domain}
\end{figure}

Fig.~\ref{single-domain} (a) shows the hysteresis loops for a thick (0.56~mg mass) Co$_3$Sn$_2$S$_2$ flake at low temperatures (80~K, 100~K, 120~K and 135~K). Here and below, the curves are obtained after zero-field cooling, the magnetic field is normal to the flake's surface,  i.e. along the Co$_3$Sn$_2$S$_2$ easy axis direction. The sample orientation is verified by the angle dependence of magnetization in Fig.~\ref{single-domain} (b) for high external magnetic field H=15~kOe.

 Hysteresis loops are of strictly rectangular shape with step-like magnetization switchings in Fig.~\ref{single-domain} (a), which confirms the single domain magnetic state of Co$_3$Sn$_2$S$_2$. The coercive field (H$_c$) is about 1~kOe, as it is expected for hard magnetic Co$_3$Sn$_2$S$_2$ material.  At the lowest 80~K temperature, the  hysteresis loop is shifted to the negative magnetic fields, as it has also been shown for Co$_3$Sn$_2$S$_2$ in Ref.~\cite{CoSnS_spin glass}.  This asymmetric shift  can be removed by multiple cycling of the external field or by temperature increase above 90~K.  The coercive field and the saturated magnetization values are monotonically diminishing with temperature in Fig.~\ref{single-domain} (a).

\begin{figure}[t]
\center{\includegraphics[width=\columnwidth]{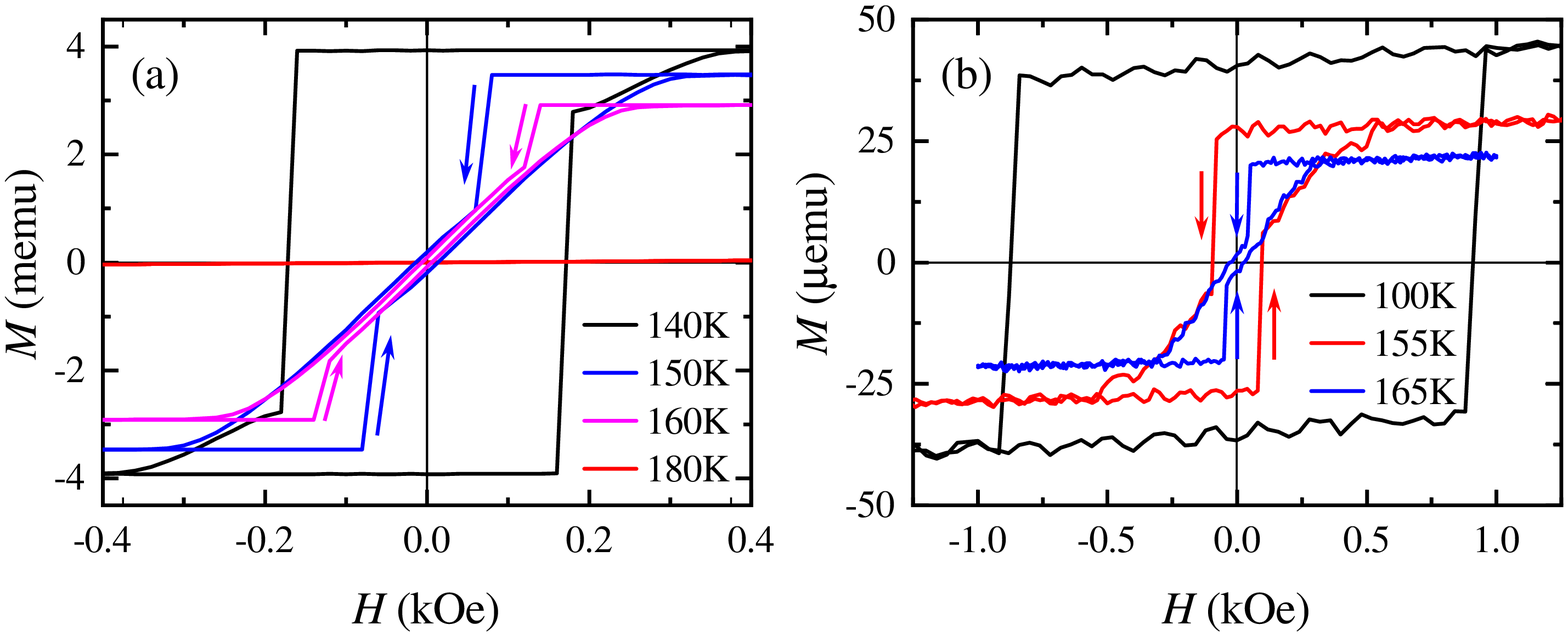}}
\caption{(Color online) (a) Hysteresis loops with step-like magnetization switchings for the  6~$\mu$m thick (0.56~mg)  Co$_3$Sn$_2$S$_2$ flake above 140~K. For every hysteresis branch, a region of slanted magnetization  appears, which accompanies the usual  step-like magnetization switching. The width of the slanted region is increasing with temperature, so the step-like magnetization switching is shifted across the zero field (the inverted hysteresis~\cite{Invh,Eb_Invh_oxfilm}. The  hysteresis is completely suppressed at 180~K, i.e. above the  Curie temperature. (b) Similar behavior for much smaller   (0.5~$\mu$m thick, 0.01~mg)  Co$_3$Sn$_2$S$_2$ single-crystal flake,  the slanted region and the inverted hysteresis appear  above 155~K in this case. The inverted hysteresis is a fingerprint of material with two independent magnetic phases~\cite{Invh}.}
\label{Tdependence}
\end{figure}

 To our surprise, the hysteresis loops show complicated temperature dependence above 140~K, see  Fig.~\ref{Tdependence} (a). For every hysteresis branch, a region of slanted magnetization dependence appears, which is not symmetric in respect to the zero magnetic field. This slanted region accompanies the usual  step-like magnetization switching of initially (at low temperatures) rectangular hysteresis loop, so there are two independent magnetization processes. First traces of the slanted region can be seen at 135~K in Fig.~\ref{single-domain} (a).  The width of the region is increasing with temperature, so the step-like magnetization switchings are shifted across the zero field. This magnetization loop type is known as the inverted hysteresis~\cite{Invh,Eb_Invh_oxfilm}. The  hysteresis is completely suppressed above the  Curie temperature, see the linear 180~K curves for two field sweep directions in Fig.~\ref{Tdependence} (a).
    
The inverted hysteresis is a fingerprint of material with two independent magnetic phases, e.g. it is realized in ferromagnetic multilayer structures~\cite{Eb_Invh_oxfilm}. In  Fig.~\ref{Tdependence} (a), one phase switches magnetization by sharp steps, while  another phase shows the slanted hysteresis loop with low coercitivity.  The inversion reflects the phase interaction in this case, so  one magnetic phase provides a bias field to the second one~\cite{Invh}. From the experiment, the phases are of strongly different temperature dependences:  the coercive field is significantly suppressed for the rectangular  loop, while the temperature-stable slanted one is of the same slope and width. Thus, the second phase can be seen only at high temperatures, cp. Figs.~\ref{single-domain} (a) and~\ref{Tdependence} (a).

Fig.~\ref{Tdependence} (b) shows qualitatively similar behavior for much smaller (0.01~mg) Co$_3$Sn$_2$S$_2$ single-crystal flake. The saturated magnetization level is of two orders of magnitude smaller in this case, but we clearly observe the slanted region and the inverted hysteresis at high temperatures, above 155~K in this case. Thus, the two-phase behavior is  universal, it can be well reproduced from sample to sample. 

FORC measurements is a standard tool to distinguish multiple magnetic phases~\cite{FORCtheory,FORCtheory1}.  Fig.~\ref{FORC} shows the initial FORC diagrams (a,c,e) and the calculated FORC-densities $\rho(H_u,H_c)$ (b,d,f) for 100~K, 135~K, 150~K temperatures, respectively.  

The multi-phase behavior can be easily seen from the raw curves even at 100~K  in Fig.~\ref{FORC} (a). For a simple rectangular hysteresis loop, one should expect the FORC diagram as two horizontal lines at positive and negative $M(H)$ saturation levels with multiple step-like switchings between them at positive magnetic fields. The multiple steps are due to the FORC procedure~\cite{FORCanalysis,Hr} with incomplete magnetization for some intermediate values of the reversal field $H_r$. This expected behavior  can indeed be seen in  Fig.~\ref{FORC} (a), but there is an additional set of FORC curves with finite slope. This type of FORC curves usually corresponds to the slanted hysteresis loop~\cite{FORCanalysis,Hr}, i.e. it indicates an additional magnetic phase even at 100~K. The first set of the curves strongly depends on temperature, see Figs.~\ref{FORC} (a,c,e), in  good correlation with the rectangular loop dependence in  Fig.~\ref{Tdependence} (a). In contrast,  the second (slanted) set of FORC curves is practically independent of temperature,  which also confirms the temperature-stable second magnetic phase. 

{
Two magnetic phases can be seen directly in the FORC-density patterns, see  Figs.~\ref{FORC} (b,d,f). For the reader's convenience, we draw the FORC density $\rho$ in the initial $(H,H_r)$ coordinates, while the $(H_u,H_c)$ axes are also shown in the plots.  For a single-phase monodomain sample, one should expect a set of $\rho(H_u,H_c)$ peaks. These peaks indeed can be seen in Figs.~\ref{FORC} (b,d,f), their $H$-positions well correspond  to the step-like switchings in the raw FORC data (a,c,e). Also, we obtain additional structures (horizontal thick green lines) in  $\rho(H,H_r)$, they are due to the slanted FORC curves with the constant slope.   Since the presence of multiple structures in the FORC density is a fingerprint of multiple magnetic phases in the material~\cite{FORCtheory,FORCtheory1},  two magnetic systems are present in the Co$_3$Sn$_2$S$_2$ single-crystal even at lowest temperatures, see especially Figs.~\ref{FORC} (a) and (b).

\begin{figure}[t]
\center{\includegraphics[width=0.9\columnwidth]{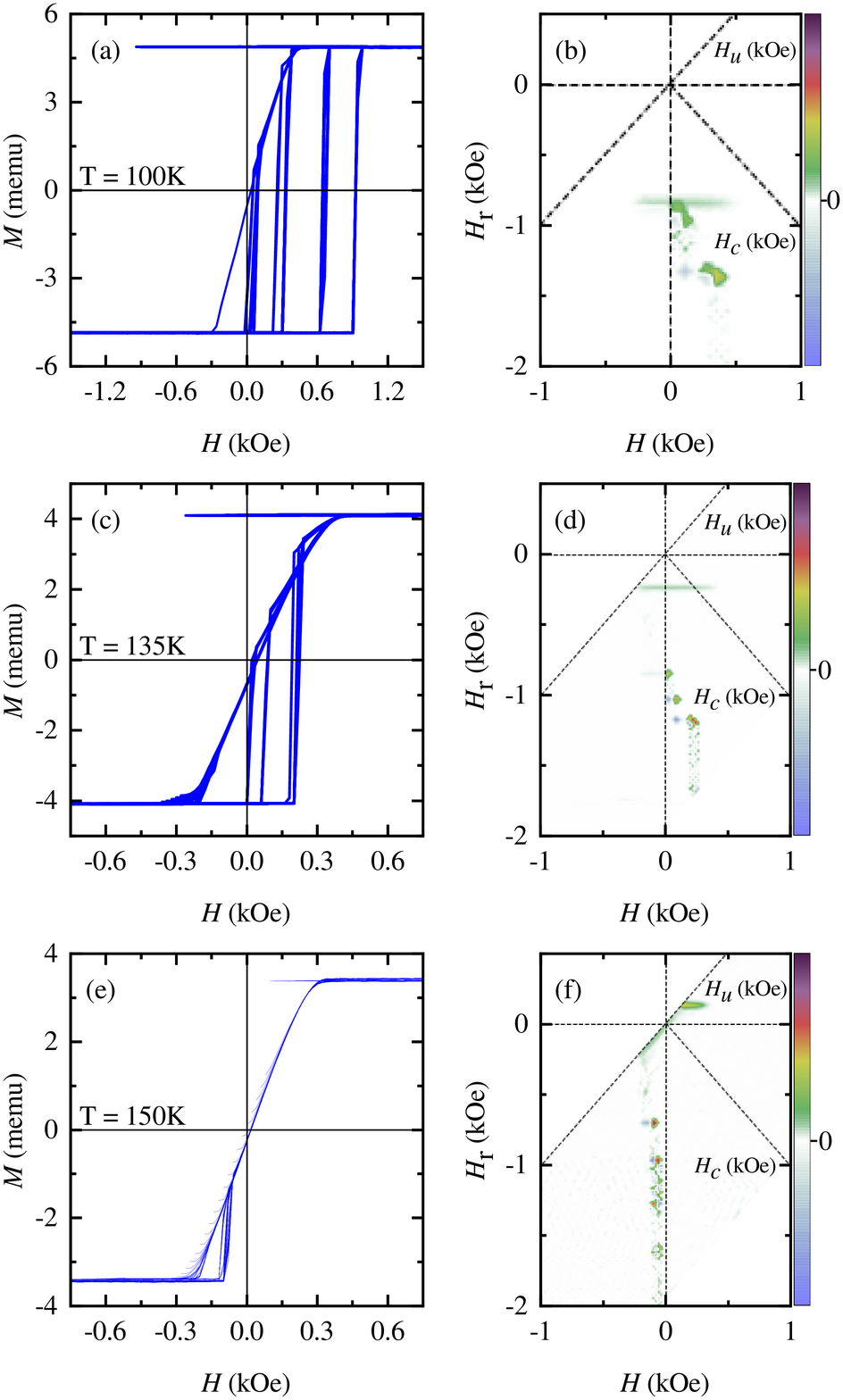}}
\caption{ (Color online) Two  magnetic phases as confirmed  by FORC measurements for the  thick (0.56~mg mass) Co$_3$Sn$_2$S$_2$ sample.  The initial FORC diagrams (a,c,e) and the calculated FORC-densities $\rho$ (b,d,f) are shown for 100~K, 135~K, 150~K temperatures, respectively.
For the reader's convenience, we draw the FORC density $\rho$ in the initial $(H,H_r)$ coordinates, while the $(H_u,H_c)$ axes are also shown in the plots.  The multi-phase behavior can be easily seen even at 100~K: in addition to multiple step-like switchings at positive magnetic fields, there is an additional set of FORC curves with the finite slope. The first set of the curves is of strong temperature dependence in (a,c,e), while the second (slanted) set of FORC curves is practically independent of temperature. In a good correspondence to the raw FORC data, we obtain an additional structure (the horizontal thick green line) in $\rho(H,H_r)$ FORC densities, see (b,d,f), which is a fingerprint of the second magnetic phase in the Co$_3$Sn$_2$S$_2$ topological semimetal.
}
\label{FORC}
\end{figure}

The described two-phase behavior is not unique for the Co$_3$Sn$_2$S$_2$ topological semimetal. Similar behavior  can also be observed for the magnetic topological nodal-line semimetal FGT, as depicted in Fig.~\ref{FGTloops} for the 0.0027~mg flake. We observe a wide slanted region in Fig.~\ref{FGTloops} (a) even at 100~K, which is accompanied by the step-like magnetization switchings, similarly to the inverted hysteresis for Co$_3$Sn$_2$S$_2$ samples above 140~K temperature in Fig.~\ref{Tdependence}, i.e. it confirms an additional magnetic phase in the sample. Two independent phases also appear in the raw FORC diagram in Fig.~\ref{FGTloops} (b) as a slanted region and several step-like switchings. The presence of multiple structures in the  FORC-density pattern is the clearest demonstration of the two-phase behavior: there are sharp peaks at negative $H_u$ values and a thick horizontal  line at the positive ones, similarly to the Co$_3$Sn$_2$S$_2$ patterns in Figs.~\ref{FORC} (b,d,f). Thus, the two-phase behavior is mostly identical for the Co$_3$Sn$_2$S$_2$ and FGT magnetic topological semimetals, only the characteristic temperatures differ for these materials.

\begin{figure}[t]
\center{\includegraphics[width=\columnwidth]{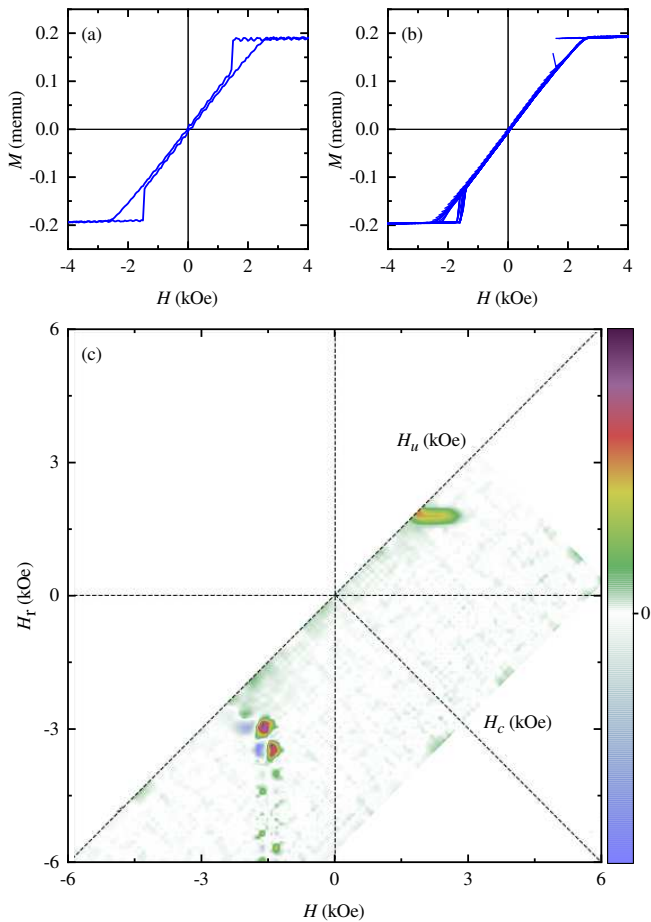}}
\caption{ (Color online)  Two-phase behavior  for the  0.45~$\mu$m thick (0.0027~mg) flake of a topological nodal-line semimetal FGT. (a) The inverted hysteresis with a wide slanted region and the step-like magnetization switchings at 100~K. Two-phase behavior is confirmed by the initial FORC diagram (b) and the calculated FORC-density $\rho$ (c). For the reader's convenience, we draw the FORC density $\rho$ in the initial $(H,H_r)$ coordinates, while the $(H_u,H_c)$ axes are also shown in the plots.  Two independent magnetic phases can be seen as multiple structures in the  FORC-density pattern: there are sharp peaks at negative $H_u$ values and a thick horizontal  line at the positive ones, so the two-phase behavior is mostly identical for  Co$_3$Sn$_2$S$_2$ and FGT magnetic topological semimetals.} 
\label{FGTloops}
\end{figure}

\section{Discussion}

Similar magnetization behavior is known for different  structures with two decoupled magnetic subsystems, e.g.  for ferromagnetic multilayers~\cite{Cu/Cofilms,GaMnAs,SrRuOheterostr} or materials with two magnetic phases~\cite{Co/CoO,nanocomposite}. Also, it has been reported for multilayer structures with topological spin textures (surface skyrmions)~\cite{Pt/Co/Ta1,Pt/Co/Ta2,Co/Pd,Ir/Fe/Co/Pt}. The inverted hysteresis corresponds to the antiferromagnetic interfacial coupling between the  magnetic phases~\cite{Exb,Invh,Eb_Invh_oxfilm}, which provides  the exchange bias field. 

We wish to mention, that the observed results can not originate from two connected flakes of Co$_3$Sn$_2$S$_2$: in the latter case, the hysteresis loop would be a sum of two rectangular ones, arbitrary shifted in magnetic field and magnetization level~\cite{GaMnAs,SrRuOheterostr}, as we indeed observe for a reference two-crystal sample. 

In principle, coexistence of two magnetic phases can been anticipated for Co$_3$Sn$_2$S$_2$ from $M(T)$ measurements in fixed magnetic fields~\cite{CoSnS_Tdepend1, CoSnS_Tdepend2, CoSnS_spin glass} and from the AHE hysteresis~\cite{CoSnS_spin glass}. While the first phase is obviously the ferromagnetic bulk, the existence of the second phase was explained by disorder effects~\cite{CoSnS_Tdepend1} or even by the spin-glass state~\cite{CoSnS_spin glass}.  

It is well known, that AHE hysteresis well correspond to the $M(H)$ magnetization reversal curves, so it is not surprising that the slanted  region appears at similar temperatures in  Fig.~\ref{Tdependence} and in Ref.~\cite{CoSnS_spin glass}. However, our FORC measurements directly demonstrate multi-phase behavior even at lowest temperatures, with excellent temperature stability of this phase in comparison with the main ferromagnetic one in Fig.~\ref{FORC}. Thus, the second phase does not appear at some specific temperature in our experiment and their robustness requires specific explanation. Also, qualitatively similar behavior is observed for Co$_3$Sn$_2$S$_2$ samples of extremely different size (two orders of magnitude in Fig.~\ref{Tdependence}), and for another magnetic topological semimetal FGT, demonstrating  universal character of the second phase.  Thus, it can hardly be ascribed to temperature-induced disorder effects~\cite{CoSnS_Tdepend1, CoSnS_Tdepend2, CoSnS_spin glass} in our experiment.

On the other hand,  Co$_3$Sn$_2$S$_2$ and FGT are magnetic semimetals with topologically protected Fermi-arc surface states~\cite{kagome_arcs,asymmr,CoSnS_Ws1,CoSnS_Ws2}. Due to the spin-momentum locking, one can expect  skyrmion-like topological spin textures on the surface of a magnetic topological semimetal. The skyrmion textures have been observed for the  Fe$_3$Sn$_2$ kagome ferromagnet~\cite{FeSn}, for FGT~\cite{FGT_skyrmion1,FGT_skyrmion2}, and recently for Co$_3$Sn$_2$S$_2$ magnetic Weyl semimetal~\cite{CoSnS_skyrmion}.

Since the surface spin textures are inherent for magnetic topological semimetals due to the spin-momentum locking in the  topological surface states, they can be regarded as the second magnetic phase. Also, the obtained hysteresis loops are of the bow-tie type in Figs.~\ref{Tdependence} and Fig.~\ref{FGTloops} (a), which is usually ascribed to skyrmions~\cite{Pt/Co/Ta,Pt/Co/Ta2,Co/Pd,Ir/Fe/Co/Pt}. The specifics of our experiment is the excellent temperature stability of the second phase in Figs.~\ref{Tdependence} and~\ref{FORC}, while the skyrmions are usually observed near the  Curie temperature. On the other hand, temperature stability is quite natural for surface-state induced spin textures due to the topological protection of surface states in topological semimetals, but  hardly can be expected for the bulk effects.

 This also explains the universal behavior of the second phase for two different topological semimetals Co$_3$Sn$_2$S$_2$ and FGT. Indeed, both these materials have different composition, structure, magnetic properties, even the bulk spectrum (Weyl one for Co$_3$Sn$_2$S$_2$, nodal-line one for FGT).  The only similarity, is the presence of the topological surface states~\cite{kagome_arcs,asymmr,CoSnS_Ws1,CoSnS_Ws2}, which, however, determines a lot of  physical properties~\cite{armitage}, e.g. magnetoresistance, anomalous Hall effect, ARPES, STM  and even an optical response. Thus, we can ascribe the second, temperature-stable magnetic phase to the surface states in topological semimetals.

\section{Conclusion}
As a conclusion, first order reversal curve analysis confirms the  two-phase behavior even at the lowest temperatures of the experiment. While the bulk ferromagnetic magnetization is expected to have strong temperature dependence, one of the observed phases demonstrates perfect stability below the  Curie temperature. Temperature stability can be expected for surface-state induced spin textures due to the topological protection of surface states in topological semimetals. The described two-phase behavior is mostly identical for Co$_3$Sn$_2$S$_2$ and Fe$_3$GeTe$_2$ magnetic topological semimetals, only the characteristic temperatures differ for    these materials. Thus,  the first order reversal curves analysis gives an evidence for  spin textures in topological semimetals even far below the Curie  temperature.

\section{Acknowledgement}

We wish to thank S.S~Khasanov for X-ray sample characterization.


\begin{thebibliography}{99}

%introduction
\bibitem{armitage} N.P.~Armitage, E.J.~Mele, and A.~Vishwanath,  Rev. Mod. Phys.  90, 015001 (2018).
\bibitem{mag1} X. Wan, A. M. Turner, A. Vishwanath, S. Y. Savrasov,  Phys. Rev. B 83, 205101
(2011).
\bibitem{mag2} M. Hirschberger, S. Kushwaha, Z. Wang, Q. Gibson, S. Liang, C. A. Belvin, B. A. Bernevig,
R. J. Cava, N. P. Ong,  Nat. Mater. 15, 1161-1165 (2016).
\bibitem{mag3} G. Xu, H. Weng, Z. Wang, X. Dai, Z. Fang,  Phys. Rev. Lett. 107, 186806 (2011).
\bibitem{mag4} S. K. Kushwaha, Z. Wang, T. Kong, R. J. Cava,  J. Phys. Condens. Matter. 30, 075701
(2018).
\bibitem{kagome}     Enke Liu, Yan Sun, Nitesh Kumar, Lukas Muechler, Aili Sun, Lin Jiao, Shuo-Ying Yang, Defa Liu, Aiji Liang, Qiunan Xu, Johannes Kroder, Vicky S\"uss, Horst Borrmann, Chandra Shekhar, Zhaosheng Wang, Chuanying Xi, Wenhong Wang, Walter Schnelle, Steffen Wirth, Yulin Chen, Sebastian T. B. Goennenwein, and Claudia Felser, Nature Physics 14, 1125 (2018)
\bibitem{kagome1} Qi Wang, Yuanfeng Xu, Rui Lou, Zhonghao Liu, Man Li, Yaobo Huang, Dawei Shen, Hongming Weng, Shancai Wang and Hechang Lei, Nature Communications 9,  3681 (2018) 
\bibitem{kagome_arcs} Noam Morali, Rajib Batabyal, Pranab Kumar Nag, Enke Liu, Qiunan Xu, Yan Sun, Binghai Yan, Claudia Felser, Nurit Avraham, Haim Beidenkopf, 	arXiv:1903.00509

\bibitem{kim} K. Kim, J. Seo, E. Lee, K.-T. Ko, B. S. Kim, Bo G. Jang, J. M. Ok, J. Lee, Y. J. Jo, W. Kang, J. H. Shim, C. Kim, H. W. Yeom, B. I. Min, B.-J. Yang, and J. S. Kim, Nat. Mater. 17, 794 (2018).
\bibitem{asymmr} S. Albarakati, C. Tan, Z. Chen, J. G. Partridge, G. Zheng, L. Farrar, E. L. H. Mayes, M. R. Field, C. Lee, Y. Wang, Y. Xiong, M. Tian, F. Xiang, A. R. Hamilton, O. A. Tretiakov, D. Culcer, Y. Zhao, and Y. Wang, Sci. Adv. 5, eaaw0409 (2019). https://doi.org/10.1126/sciadv.aaw0409


\bibitem{jiang15} J.~Jiang, F.~Tang, X.C.~Pan, H.M.~Liu, X.H.~Niu, Y.X.~Wang, D.F.~Xu, H.F.~Yang, B.P.~Xie, F.Q.~Song, P.~Dudin, T.K.~Kim, M.~Hoesch, P.K.~Das, I.~Vobornik, X.G.~Wan, and D.L.~Feng,
Phys. Rev. Lett. 115, 166601 (2015).
\bibitem{rhodes15} D.~Rhodes, S.~Das, Q.R.~Zhang, B.~Zeng, N.R.~Pradhan, N.~Kikugawa, E.~Manousakis, and L.~Balicas, Phys. Rev. B 92, 125152 (2015).
\bibitem{wang16} Y.Wang, K.Wang, J. Reutt-Robey, J. Paglione, and M. S. Fuhrer,
Phys. Rev. B 93, 121108 (2016).

\bibitem{Sp-m-lock} Su-Yang Xu, Chang Liu, Satya K. Kushwaha, Raman Sankar, Jason W. Krizan, Ilya Belopolski, Madhab Neupane, Guang Bian, Nasser Alidoust, Tay-Rong Chang, Horng-Tay Jeng, Cheng-Yi Huang, Wei-Feng Tsai, Hsin Lin, Pavel P. Shibayev, Fang-Cheng Chou, Robert J. Cava, and M. Zahid Hasan,
Science, 347 (6219), DOI: 10.1126/science.1256742.

\bibitem{das16} P.K.~Das, D.D.~Sante, I.~Vobornik, J.~Fujii, T.~Okuda, E.~Bruyer, A.~Gyenis, B.E.~Feldman, J.~Tao, R.~Ciancio, G.~Rossi, M.N.~Ali, S.~Picozzi, A.~Yadzani, G.~Panaccione, and R.J.~Cava, Nature
Comm. 7, 10847 (2016).
\bibitem{feng2016} B.~Feng, Y.-H.~Chan, Y.~Feng, R.-Y.~Liu,1 M.-Y.~Chou, K.~Kuroda, K.~Yaji, A.~Harasawa,
P.~Moras, A.~Barinov, W.~Malaeb, C.~Bareille, T.~Kondo, S.~Shin, F.~Komori, T.-C.~Chiang, Y.~Shi, and I.~Matsuda, Phys Rev B 94, 195134 (2016).
\bibitem{lv2015} B.Q.~Lv, S.~Muff, T.~Qian, Z.D.~Song, S.M.~Nie, N.~Xu, P.~Richard, C.E.~Matt, N.C.~Plumb, L.X.~Zhao, G.F.~Chen, Z.~Fang, X.~Dai, J.H.~Dil, J.~Mesot, M.~Shi, H.M.~Weng, and H.~Ding, Phys. Rev. Lett. 115, 217601 (2015).
\bibitem{xu16} S.-Y.~Xu, I.~Belopolski, D.S.~Sanchez, M.~Neupane, G.~Chang, K.~Yaji, Z.~Yuan, C.~Zhang, K.~Kuroda, G.~Bian, C.~Guo, H.~Lu, T.-R.~Chang, N.~Alidoust, H.~Zheng, C.-C.~Lee, S.-M.~Huang, C.-H.~Hsu, H.-T.~Jeng, A.~Bansil, T.~Neupert, F.~Komori, T.~Kondo, S.~Shin, H.~Lin, S.~Jia, and M.Z.~Hasan, Phys. Rev. Lett. 116, 096801 (2016).


\bibitem{CrGeTe} Myung-Geun Han, Joseph A. Garlow, Yu Liu, Huiqin Zhang, Jun Li, Donald DiMarzio, Mark W. Knight, Cedomir Petrovic, Deep Jariwala and Yimei Zhu, Nano Lett., 19, 11, 7859–7865 (2019).
\bibitem{FGT_skyrmion1} Bei Ding, Zefang Li, Guizhou Xu, Hang Li, Zhipeng Hou, Enke Liu, Xuekui Xi, Feng Xu, Yuan Yao, and Wenhong Wang, Nano Lett., 20, 868--873 (2020).
\bibitem{FGT_skyrmion2} Giang D. Nguyen, Jinhwan Lee, Tom Berlijn, Qiang Zou, Saban M. Hus, Jewook Park, Zheng Gai, Changgu Lee, and An-Ping Li, Physical Review B 97, 014425 (2018).
\bibitem{Araki} Yasufumi Araki, Ann. Phys. (Berlin), 532, 1900287, 1--16 (2020). 




\bibitem{Co/Pd} Robert Streubel, Luyang Han, Mi-Young Im, Florian Kronast, Ulrich K. R$\ddot{o}$ler, Florin Radu, Radu Abrudan, Gungun Lin, Oliver G. Schmidt, Peter Fischer and Denys Makarov, Scientific Reports, 5, 8787 (2015).
\bibitem{Pt/Co/Ta} Senfu Zhang, Junwei Zhang, Yan Wen, Eugene M. Chudnovsky and Xixiang Zhang, Communications Physics, 1, 36 (2018).
\bibitem{Pt/Co/Ta1} Senfu Zhang, Junwei Zhang, Qiang Zhang, Craig Barton, Volker Neu, Yuelei Zhao, Zhipeng Hou, Yan Wen, Chen Gong, Olga Kazakova, Wenhong Wang, Yong Peng, Dmitry A. Garanin, Eugene M. Chudnovsky and Xixiang Zhang, Applied Physics Letters 112, 132405 (2018).
\bibitem{Pt/Co/Ta2} You Ba, Shihao Zhuang, Yike Zhang, Yutong Wang, Yang Gao, Hengan Zhou, Mingfeng Chen, Weideng Sun, Quan Liu, Guozhi Chai, Jing Ma, Ying Zhang, Huanfang Tian, Haifeng Du, Wanjun Jiang, Cewen Nan, Jia-Mian Hu and Yonggang Zhao, Nature Communications, 12, 322 (2021); https://doi.org/10.1038/s41467-020-20528-y.


\bibitem{Ir/Fe/Co/Pt} Anjan Soumyanarayanan, M. Raju, A. L. Gonzalez Oyarce, Anthony K. C. Tan, Mi-Young Im, A. P. Petrovi\'c, Pin Ho, K. H. Khoo, M. Tran, C. K. Gan, F. Ernult and C. Panagopoulos, Nature Mater, 16, 898–904 (2017). https://doi.org/10.1038/nmat4934.

\bibitem{SrRuOheterostr}  Lena Wysocki,  Sven Erik Ilse,  Lin Yang, Eberhard Goering,  Felix Gunkel,  Regina Dittmann,  Paul H. M. van Loosdrecht and  Ionela Lindfors-Vrejoiu, Journal of Applied Physics 131, 133902 (2022); https://doi.org/10.1063/5.0087098.
\bibitem{invhyst} M. Charilaou, C. Bordel, and F. Hellman, Appl. Phys. Lett. 104, 212405 (2014).
\bibitem{exchbias} S. Maat, K. Takano, S.S.P. Parkin, and Eric E. Fullerton, Physical Review Letters, 87, 8 (2001).

\bibitem{spin valve1} B. Dieny, V. S. Speriosu, S. S. P. Parkin, B. A. Gurney, D. R. Wilhoit, and D. Mauri, Physical Review B, 43, 1297-1300 (1991).
\bibitem{spin valve2} R. Q. Zhang, J. Su, J. W. Cai, G. Y. Shi, F. Li, L. Y. Liao, F. Pan, and C. Song, Appl. Phys. Lett. 114, 092404 (2019).



\bibitem{bow-tie} Felipe Tejo, Denilson Toneto, Sim\'on Oyarz\'un, Jos\'e Hermosilla, Caroline S. Danna, Juan L. Palma, Ricardo B. da Silva, Lucio S. Dorneles, and Juliano C. Denardin, ACS Appl. Mater. Interfaces, 12, 47, 53454 (2020).



\bibitem{timnal} V. D. Esin, D. N. Borisenko, A. V. Timonina, N. N. Kolesnikov, and E. V. Deviatov
	Phys. Rev. B 101, 155309 (2020)
	DOI:https://doi.org/10.1103/PhysRevB.101.155309
\bibitem{wteni}	A. Kononov, O. O. Shvetsov, A. V. Timonina, N. N. Kolesnikov, E. V. Deviatov
	JETP Letters, 109, 180 (2019)
	DOI: 10.1134/S0021364019030020
\bibitem{cosns}	O. O. Shvetsov, V. D. Esin, A. V. Timonina, N. N. Kolesnikov, E. V. Deviatov
	EPL, 127, 57002 (2019)
	doi: 10.1209/0295-5075/127/57002
	
\bibitem{myers} E.B. Myers, D.C. Ralph, J.A. Katine, R.N. Louie, R.A. Buhrman, Science, 285, 867 (1999).
\bibitem{tsoi1} M. Tsoi, A. G. M. Jansen, J. Bass, W.-C. Chiang, M. Seck, V. Tsoi, and P. Wyder, Phys. Rev. Lett., 80, 4281 (1998).
\bibitem{tsoi2} M. Tsoi, A. G. M. Jansen, J. Bass, W.-C. Chiang, V. Tsoi  and P. Wyder, Nature, 406, 46, (2000).
\bibitem{katine} J. A. Katine, F. J. Albert, R. A. Buhrman, E. B. Myers and D. C. Ralph,  Phys. Rev. Lett., 84, 3149 (2000)

%samples

\bibitem{fgtoleg} O. O. Shvetsov, Yu. S. Barash, A. V. Timonina, N. N. Kolesnikov, E. V. Deviatov, JETP Letters, 115, 267 (2022). 
	DOI: 10.1134/S0021364022100101
	
\bibitem{ahe1} For a review on AHE, see N. Nagaosa, J. Sinova, S. Onoda, A.H. MacDonald, and P.P. Ong,  Rev. Mod. Phys. 82, 1539 (2010).
\bibitem{ahe2}	F. D. M.  Haldane,  Phys. Rev. Lett. 93, 206602 (2004).
\bibitem{ahe3} S. Nakatsuji, N. Kiyohara,and T. Higo,  Nature 527, 212 (2015).

\bibitem{Tc} W. Schnelle, A. Leithe-Jasper, H. Rosner, F. M. Schappacher, R. P$\ddot{o}$ttgen, F. Pielnhofer and R. Weihrich,  Physical Review B, 88, 144404 (2013).



\bibitem{FORCanalysis} B. C. Dodrill, Magnetometry Measurements and First-Order-Reversal-Curve (FORC) 
Analysis, Lake Shore Cryotronics. www.lakeshore.com.
\bibitem{Hr} Dustin A. Gilbert, Peyton D. Murray, Julius De Rojas, Randy K. Dumas, Joseph E. Davies and Kai Liu, Scientific Reports, 11, 4018 (2021).
\bibitem{FORCtheory} B. C. Dodrill, H. S. Reichard, and T. Shimizu, Lake Shore Cryotronics. Technical Note. www.lakeshore.com.
\bibitem{FORCtheory1} B. C. Dodrill, Magnetometry Measurements of Nanomagnetic Materials, Advanced Materials: ThechConnect Briefs 2018 www.lakeshore.com. 

%results 


\bibitem{CoSnS_spin glass} Ella Lachman, Ryan A. Murphy, Nikola Maksimovic, Robert Kealhofer, Shannon Haley, Ross D. McDonald, Jeffrey R. Long and James G. Analytis, Nature Communications, 11,  560 (2020).

\bibitem{Invh} M. J. O'Shea and A.-L. Al-Sharif, Journal of Applied Physics 75, 6673 (1994).
\bibitem{Eb_Invh_oxfilm} Mohammad Saghayezhian, Zhen Wang, Hangwen Guo, Rongying Jin, Yimei Zhu, Jiandi Zhang and E. W. Plummer Physical Review Research 1, 033160 (2019).

 



%discussion


\bibitem{Cu/Cofilms} T. R. McGuire and T. S. Plaskett, IEEE Transactions on Magnetics,  28, 2748-2750 (1992).
\bibitem{GaMnAs} S. Mark, C. Gould, K. Pappert, J. Wenisch, K. Brunner, G. Schmidt, and L. W. Molenkamp, Physical Review Letters, 103, 017204 (2009) 
\bibitem{Co/CoO} Jes\'us G. Ovejero, Vanda Godinho, Bertrand Lacroix, Miguel A. Garc\'ia, Antonio Hernando, Asunci\'on Fern\'andez, Materials and Design, 171, 107691 (2019), https://doi.org/10.1016/j.matdes.2019.107691.
\bibitem{nanocomposite} Zhao-hua Cheng, Jun-xian Zhang and H. Kronm$\ddot{u}$ller, Physical Review B,  68, 144417 (2003). 

\bibitem{Exb}J. Nogu\'es, Ivan K. Schuller, Journal of Magnetism and Magnetic Materials, 192, 203 (1999).

\bibitem{CoSnS_Tdepend1} Z. Guguchia, J. A. T. Verezhak, D. J. Gawryluk, S. S. Tsirkin, J.-X. Yin, I. Belopolski, H. Zhou, G. Simutis, S.-S. Zhang, T. A. Cochran, G. Chang, E. Pomjakushina, L. Keller, Z. Skrzeczkowska, Q. Wang, H. C. Lei, R. Khasanov, A. Amato, S. Jia, T. Neupert, H. Luetkens and M. Z. Hasan, Nature Communications, 11, 559 (2020).
\bibitem{CoSnS_Tdepend2} H.C. Wu, P.J. Sun, D.J. Hsieh, H.J. Chen, D. Chandrasekhar Kakarla, L.Z. Deng,
C.W. Chu, H.D. Yang, Materials Today Physics, 12, 100189 (2020).


\bibitem{CoSnS_Ws1} Noam Morali, Rajib Batabyal, Pranab Kumar Nag, Enke Liu, Qiunan Xu,Yan Sun, Binghai Yan, Claudia Felser, Nurit Avraham, Haim Beidenkopf, Science 365, 1286 (2019).
\bibitem{CoSnS_Ws2} Qi Wang, Yuanfeng Xu, Rui Lou, Zhonghao Liu, Man Li, Yaobo Huang, Dawei Shen, Hongming Weng, Shancai Wang and Hechang Lei, Nature Communications, 9, 3681 (2018).
\bibitem{FeSn} Zhipeng Hou, Weijun Ren, Bei Ding, Guizhou Xu, Yue Wang, Bing Yang, Qiang Zhang, Ying Zhang, Enke Liu, Feng Xu, Wenhong Wang, Guangheng Wu, Xixiang Zhang, Baogen Shen, Zhidong Zhang, Advanced Materials, 30,  1701144 (2017).

\bibitem{CoSnS_skyrmion} Akira Sugawara, Tetsuya Akashi, Mohamed A. Kassem, Yoshikazu Tabata, Takeshi Waki, and Hiroyuki Nakamura, Physical Review Materials 3, 104421 (2019).






 





\end{thebibliography}
\end{document}